\documentclass[letter,11pt]{article}
\usepackage{jheppub}

\def\as{\alpha_S}
\def\eps{\varepsilon}
\def\lrhosq{\ln^2(\rho)}
\def\lrho{\ln(\rho)}

\def\lbe{\ln(\beta)}

\title{\boldmath NNLO corrections to top-pair production at hadron colliders: the all-fermionic scattering channels}

\author[a]{Micha\l{}  Czakon}
\author[b]{and Alexander Mitov}

\affiliation[a]{Institut f\"ur Theoretische Teilchenphysik und Kosmologie,
RWTH Aachen University, D-52056 Aachen, Germany}
\affiliation[b]{Theory Division, CERN, CH-1211 Geneva 23, Switzerland}

\note{Preprint numbers: CERN-PH-TH/2012-181, TTK-12-30}

\abstract{This is a second paper in our ongoing calculation of the next--to--next--to--leading order (NNLO) QCD correction to the total inclusive top--pair production cross-section at hadron colliders. In this paper we calculate the reaction $q\bar q \to t\bar t + q\bar q$ which was not considered in our previous work on $q\bar q \to t\bar t +X$ \cite{Baernreuther:2012ws} due to its phenomenologically negligible size. We also calculate all remaining fermion--pair--initiated partonic channels $qq',~ q\bar q'$ and $qq$ that contribute to top-pair production starting from NNLO.  The contributions of these reactions to the total cross-section for top-pair production at the Tevatron and LHC are small, at the permil level. The most interesting feature of these reactions is their characteristic logarithmic rise in the high energy limit. We compute the constant term in the leading power behavior in this limit, and achieve precision that is an order of magnitude better than the precision of a recent theoretical prediction for this constant. All four partonic reactions computed in this paper are included in our numerical program {\tt Top++}. The calculation of the NNLO corrections to the two remaining partonic reactions, $qg\to t\bar t+X$ and $gg\to t\bar t+X$, is ongoing.}

\begin{document} 
\maketitle
\flushbottom

\section{Introduction}

Until very recently top-pair production at hadron colliders was analyzed in improved next-to-leading order (NLO) QCD. Broadly speaking, one can identify two such approaches, usually referred to in the literature as {\it resummed} and {\it approximate NNLO}, the latter simply being the truncation of the former to order ${\cal O}(\as^4)$. 

The improved--NLO approximation to the NNLO cross-section is based on the next--to--next--to--leading log (NNLL) threshold approximation \cite{Beneke:2009rj,Czakon:2009zw} and also includes Coulombic terms \cite{Beneke:2009ye} through NNLO. This approach is valid close to absolute threshold and the approximate results are added to the well known NLO \cite{Nason:1987xz,Beenakker:1988bq,Czakon:2008ii} and NLL \cite{Bonciani:1998vc} results. Alternatively, in Refs.~\cite{Ahrens:2010zv,Ahrens:2011mw,Ahrens:2011px} the resummed NNLL (and, by truncation, the approximate NNLO) total inclusive cross-section was derived from the resummed differential one. 

A number of phenomenological studies have been presented in the literature \cite{Ahrens:2010zv,Ahrens:2011mw,Ahrens:2011px,arXiv:0906.5273,Beneke:2010fm,Kidonakis:2010dk,Kidonakis:2011ca,Beneke:2011mq,Cacciari:2011hy}. Critical comparisons of the various approaches can be found in Refs.~\cite{Kidonakis:2011ca,Cacciari:2011hy,Beneke:2011ys}.

As was demonstrated in Ref.~\cite{Cacciari:2011hy}, a true improvement in the theoretical precision in top-pair production at both Tevatron and LHC can be expected only upon inclusion of the full NNLO correction to the partonic cross-section. In our recent paper \cite{Baernreuther:2012ws} we computed the dominant correction to top-pair production at the Tevatron confirming the expectations set in Ref.~\cite{Cacciari:2011hy}. 

At present theory agrees with data from the Tevatron and LHC~\cite{Abazov:2011cq,Abazov:2011mi,Aaltonen:2011tm,Aaltonen:2010bs,Khachatryan:2010ez,Chatrchyan:2011nb,Aad:2010ey,:1900yx,Aad:2011yb,CDF-top-sigma,ATLAS-top-sigma,CMS-top-sigma} and the NNLO theoretical prediction for the Tevatron \cite{Baernreuther:2012ws} has significantly smaller uncertainty than the existing experimental measurements. Theoretical predictions \cite{Brodsky:2012sz} based on the so-called BLM/PMC approach \cite{Brodsky:1982gc,Brodsky:2011ig,Brodsky:2011ta} have recently appeared and they, too, exhibit very small theoretical uncertainty. A detailed comparison with the exact NNLO result is of interest and will be performed elsewhere.

In Ref.~\cite{Baernreuther:2012ws} we argued that the purely fermionic channels that were not included there were phenomenologically negligible. Nevertheless, these reactions exhibit a logarithmic rise at high energy which could make them relevant for the description of lighter quarks, like {\it charm} or {\it bottom}, or for top-pair production at future higher energy hadron colliders. To that end, in this work we complete the calculation of all fermion-pair initiated contributions to top-pair production at hadron colliders through NNLO. 
Specifically, we calculate the reactions $qq\to t\bar t + X$,  $qq'\to t\bar t + X$ and $q\bar q'\to t\bar t + X$, with $q'\neq q$. We also complete the computation of the $q\bar q$ initiated reaction \cite{Baernreuther:2012ws} by deriving the result for the numerically subdominant reaction $q\bar q\to t\bar t + q\bar q$ which was not considered in Ref.~\cite{Baernreuther:2012ws}. The results derived in this paper fully confirm our expectations \cite{Baernreuther:2012ws} about the size of these reactions. Within the numerical accuracy of our calculation, we confirm the exactly predicted \cite{Ball:2001pq} leading high-energy logarithmic term of these reactions. Moreover, we are able to extract the subleading constant term in the high-energy expansion of the cross-sections. Our result is consistent with a very recent prediction \cite{Moch:2012mk} for this constant and improves the precision with which this constant is known by one order of magnitude.

The paper is organized as follows: in section \ref{sec:notation} we specify the reactions that we compute in the present work. Our calculational approach is explained in section \ref{sec:calc}. Factorization of collinear singularities is worked out in section \ref{sec:factorization}, while in section \ref{sec:scale} the scale dependence is derived. The explicit results are given in section \ref{sec:results}. We conclude with section \ref{sec:pheno}, where we discuss the phenomenological significance of the results computed in this work.

\section{Notation}\label{sec:notation}

Following the notation of Ref.~\cite{Baernreuther:2012ws}, the total inclusive top-pair production cross-section reads:
\begin{equation}
\sigma_{\rm tot} = \sum_{i,j} \int_0^{\beta_{\rm max}}d\beta\, \Phi_{ij}(\beta,\mu^2)\, \hat\sigma_{ij}(\beta,m^2,\mu^2)  \, ,
\label{eq:sigmatot}
\end{equation}
where $i,j$ run over all possible initial sate partons, $\beta_{\rm max} \equiv \sqrt{1-4m^2/S}$ with $\sqrt{S}$ the c.m. energy of the hadron collider and $\beta=\sqrt{1-4m^2/s}$ is the relative velocity of the final state top quarks having pole mass $m$ and produced at partonic c.m. energy $\sqrt{s}$. 

The partonic flux appearing in Eq.~(\ref{eq:sigmatot}) reads 
\begin{equation}
\Phi_{ij}(\beta,\mu^2) = {2\beta \over 1-\beta^2}~ {\cal L}_{ij}\left({1-\beta_{\rm max}^2\over 1-\beta^2}, \mu^2\right) \, ,
\label{eq:flux}
\end{equation}
where, as usual, the partonic luminosity is 
\begin{equation}
{\cal L}_{ij}(x,\mu^2) = x \left( f_i\otimes f_j \right) (x,\mu^2) = x \int_0^1 dy \int_0^1 dz \, \delta(x-yz) f_i(y)f_j(z) \, .
\label{eq:Luminosity}
\end{equation}

The scale $\mu$ in Eq.~(\ref{eq:sigmatot}) stands for both the renormalization ($\mu_R$) and factorization scales ($\mu_F$). For $\mu_F=\mu_R=\mu$ the NNLO partonic cross-section for the reaction $ij\to t\bar t + X$ reads
\begin{eqnarray}
\hat\sigma_{ij}\left(\beta,m^2,\mu^2\right) = {\as^2\over m^2}\Bigg\{  \sigma^{(0)}_{ij} + \as \left[ \sigma^{(1)}_{ij} + L\, \sigma^{(1,1)}_{ij} \right] + \as^2\left[ \sigma^{(2)}_{ij} + L\, \sigma^{(2,1)}_{ij} + L^2 \sigma^{(2,2)}_{ij} \right] \Bigg\} \, ,
\label{eq:sigmapart}
\end{eqnarray}
where $L = \ln\left(\mu^2/m^2\right)$ and $\as$ is the ${\overline {\rm MS}}$ coupling renormalized with $N_L=5$ active flavors at scale $\mu^2$. The functions $\sigma^{(n(,m))}_{ij}$ depend only on $\beta$.

The partonic cross-sections are known exactly \cite{Nason:1987xz,Beenakker:1988bq,Czakon:2008ii} through NLO. The scale controlling functions $\sigma^{(2,1)}_{ij}$ and $\sigma^{(2,2)}_{ij}$ can be easily computed from the NLO results $\sigma^{(1)}_{ij}$, see section \ref{sec:scale}. The dependence on $\mu_R\neq \mu_F$ can be trivially restored in Eq.~(\ref{eq:sigmapart}) by re-expressing $\as(\mu_F)$ in powers of $\as(\mu_R)$; see for example Ref.~\cite{arXiv:0906.5273}. 

The dominant, phenomenologically relevant part of the reaction $q\bar q\to t\bar t + X$ was computed through NNLO in Ref.~\cite{Baernreuther:2012ws}. In this paper we compute the NNLO corrections to the remaining part of this reaction as well as the three new fermionic reactions $qq\to t\bar t + X,~qq'\to t\bar t + X,~q\bar q'\to t\bar t + X$, with $q'\neq q$: 
\begin{eqnarray}
q\bar q &\to& t\bar t + q\bar q\big\vert_{\rm NS} \, , \label{eq:qqbarID}\\
q\bar q' &\to& t\bar t + q\bar q' \, , \label{eq:qqbar'}\\
q q' &\to& t\bar t + q q' \, , \label{eq:qq'}\\
q q &\to& t\bar t + q q \, . \label{eq:qq}
\end{eqnarray}

The label $NS$ in Eq.~(\ref{eq:qqbarID}) implies that we consider only those squared diagrams where the $q\bar q$ final state is not produced by a pure gluon splitting. Such (singlet) contributions are equal for all massless flavors, contribute with a power of $N_L$, and have been included in the calculation of Ref.~\cite{Baernreuther:2012ws}.

The currently unknown contributions to $t\bar t$ production at NNLO are the $qg$ and $gg$ initiated reactions. They will be the subject of a future publication.

\section{The calculation}\label{sec:calc}

All four partonic reactions (\ref{eq:qqbarID},\ref{eq:qqbar'},\ref{eq:qq'},\ref{eq:qq}) considered in this paper are of the so-called Double-Real type, i.e. they all have four-particle final states. They are computed with the STRIPPER approach of Refs.~\cite{Czakon:2010td,Czakon:2011ve}. We organize the calculation of the bare diagrams in the following way: we directly compute the bare contributions
\begin{eqnarray}
\tilde\sigma_\eps(q\bar q' \to t\bar t + q\bar q')~~{\rm and}~~ \tilde\sigma_\eps(qq' \to t\bar t + qq') 
\label{eq:primereactions}
\end{eqnarray}
to the reactions (\ref{eq:qqbar'},\ref{eq:qq'}) while the two remaining partonic reactions (\ref{eq:qqbarID},\ref{eq:qq}) are computed as differences with respect to the reactions (\ref{eq:qqbar'},\ref{eq:qq'})
\begin{eqnarray}
\sigma^{\rm diff}_\eps(q\bar q) &=&\tilde\sigma_\eps(q\bar q \to t\bar t + q\bar q\big\vert_{\rm NS}) - \tilde\sigma_\eps(q\bar q' \to t\bar t + q\bar q')  \, , \label{eq:sigmadiffqqbar}\\
\sigma^{\rm diff}_\eps(qq) &=&\tilde\sigma_\eps(qq \to t\bar t + qq) - \tilde\sigma_\eps(qq' \to t\bar t + qq')  \, . \label{eq:sigmadiffqq}
\end{eqnarray}
The differences $\sigma^{\rm diff}_\eps$ are derived from pure interference diagrams and vanish both at threshold $\beta = 0$ and in the high-energy limit $\beta = 1$.

The subscript $\eps$ appearing in Eqs.~(\ref{eq:primereactions},\ref{eq:sigmadiffqqbar},\ref{eq:sigmadiffqq}) emphasizes that these are bare cross-sections, containing collinear singularities starting from $1/\eps^2$. To subtract these singularities and obtain the finite partonic cross-sections $\hat\sigma$, one needs to perform collinear factorization, which we describe next.

\section{Collinear factorization}\label{sec:factorization}

The description of the collinear factorization deserves some attention since for the reactions considered in this paper it has not been spelled out in the literature. Moreover, the collinear factorization for the reaction (\ref{eq:qqbarID}) represents a nonstandard contribution to the reaction $q\bar q\to t\bar t + X$ and, for consistency, was suppressed in Ref.~\cite{Baernreuther:2012ws}. We take the opportunity to describe it in this work.

In the notation of Eq.~(\ref{eq:sigmapart}), and setting $\mu=m$, the bare partonic cross-sections read
\begin{equation}
\tilde\sigma_{ij}(\epsilon, \rho) = {\as^2\over m^2}\left\{\tilde\sigma_{ij}^{(0)}(\epsilon, \rho) + \as\tilde\sigma_{ij}^{(1)}(\epsilon, \rho) + \as^2\tilde\sigma_{ij}^{(2)}(\epsilon, \rho) +\dots \right\} \, .
\label{eq:sigma-def}
\end{equation}
They are defined in $d=4-2\eps$ dimensions and expressed in terms of the dimensionless variable $\rho=4m^2/s=1-\beta^2$. To obtain the finite ${\overline {\rm MS}}$-subtracted partonic cross-sections $\hat\sigma_{ij}(\rho)$ one has to factor out the initial state collinear singularities:
\footnote{We note a typo in Eq.(7) of Ref.~\cite{Czakon:2008ii}, where $\sigma$ and $\hat\sigma$ have been exchanged. This typo does not affect the rest of Ref.~\cite{Czakon:2008ii}.}
\begin{equation}
{\tilde\sigma_{ij}(\epsilon, \rho)\over \rho} = \sum_{k,l}\left[ {\hat\sigma_{kl}(x)\over x} \otimes \Gamma_{ki} \otimes \Gamma_{lj}\right](\rho) \, .
\label{eq:sigma-fact}
\end{equation}

The ${\overline {\rm MS}}$ collinear counterterms $\Gamma$ are expressed through the space-like splitting functions $P^{(n)}_{ij}$, defined as an expansion in $(\as/(2\pi))^n$. Through NNLO we have:
\begin{eqnarray}
\label{Gamma}
\Gamma_{ij}(\epsilon,x) &=& \delta_{ij}\delta(1-x) + \as\Gamma^{(1)}_{ij}(\epsilon,x) + \as^2\Gamma^{(2)}_{ij}(\epsilon,x)  \, ,\\
\Gamma^{(1)}_{ij}(\epsilon,x) &=& -{1\over 2\pi} ~{P^{(0)}_{ij}(x)\over \epsilon} \, , \nonumber \\
\Gamma^{(2)}_{ij}(\epsilon,x) &=& \left({1\over 2\pi}\right)^2  \Bigg\{ {1\over 2\epsilon^2}\left[P^{(0)}_{ik} \otimes P^{(0)}_{kj}(x) + \beta_0P^{(0)}_{ij}(x)\right] - {1\over 2\epsilon} P^{(1)}_{ij}(x) \Bigg\}\, , \nonumber
\end{eqnarray}
with $\beta_0=11C_A/6-N_L/3$ and $\as$ the renormalized coupling at scale $\mu_R$. 
 
It is more convenient to introduce the functions $\tilde s^{(n)}_{ij}$ and $s^{(n)}_{ij}$ defined as $\tilde s_{ij}^{(n)}(\eps,\rho) \equiv \tilde\sigma^{(n)}_{ij}(\eps,\rho)/\rho$ and $s^{(n)}_{ij}(\rho) \equiv \sigma^{(n)}_{ij}(\rho)/\rho$. In terms of these functions the finite cross-sections read:
\begin{eqnarray}
s^{(0)}_{ij} &=& \tilde s^{(0)}_{ij} \, ,\\
s^{(1)}_{ij} &=& \tilde s^{(1)}_{ij} - \Gamma^{(1)}_{ki}\otimes \tilde s^{(0)}_{kj} - \tilde s^{(0)}_{ik} \otimes \Gamma^{(1)}_{kj}\, ,  \nonumber\\
s^{(2)}_{ij} &=& \tilde s^{(2)}_{ij} - \left(\Gamma^{(2)}_{ki}-\Gamma^{(1)}_{kl}\otimes \Gamma^{(1)}_{li} \right)\otimes \tilde s^{(0)}_{kj}- \tilde s^{(0)}_{ik} \otimes \left(\Gamma^{(2)}_{kj}-\Gamma^{(1)}_{kl}\otimes \Gamma^{(1)}_{lj} \right) + \Gamma^{(1)}_{ki}\otimes \tilde s^{(0)}_{kl} \otimes \Gamma^{(1)}_{lj}\nonumber\\
&& - \tilde s^{(1)}_{ik} \otimes \Gamma^{(1)}_{kj} - \Gamma^{(1)}_{ki}\otimes \tilde s^{(1)}_{kj} \, .\nonumber
\end{eqnarray}
For brevity, above we have suppressed the dependence on $\epsilon$ and $x$.

Next we consider the $qq,~qq'$ and $q\bar q'$ initiated reactions with $q'\neq q$. Introducing the notation $\tilde q =(q,q',\bar q')$, we get:
\begin{eqnarray}
s^{(2)}_{q \tilde q} &=& \tilde s^{(2)}_{q \tilde q} +  \left({1\over 2\pi}\right)^2 \tilde s^{(0)}_{q \bar q} \otimes \Bigg\{ {1\over \epsilon^2} P^{(0)}_{qg}\otimes P^{(0)}_{gq} + {1\over \epsilon} P^{(1)}_{q\overline{\tilde q}} \Bigg\} \nonumber\\
&& + {1\over \epsilon^2} \left({1\over 2\pi}\right)^2 \tilde s^{(0)}_{gg} \otimes P^{(0)}_{gq}\otimes P^{(0)}_{gq} + {2\over \epsilon} \left({1\over 2\pi}\right) \tilde s^{(1)}_{gq} \otimes P^{(0)}_{gq} \, .
\label{shat2}
\end{eqnarray}
The function $P^{(1)}_{q\overline{\tilde q}}$ reads:
\begin{equation}
P^{(1)}_{q\overline{\tilde q}} = \left\{ 
  \begin{array}{l l}
    P^{(1),S}_{qq} +\, P^{(1),V}_{q\bar q} & ~ {\rm if}\quad \tilde q = q\, , \\
    P^{(1),S}_{qq} & ~ {\rm if}\quad \tilde q = (q',\bar q')\, ,
  \end{array} \right.
\label{P1qq}
\end{equation}
where $P^{(1),S},~P^{(1),V}$ are the singlet and ($q\bar q$) valence NLO splitting functions in the notation of Ref.~\cite{ESW}. The NLO functions $\tilde s^{(1)}_{ij}$ appearing in this section are needed through order ${\cal O}(\eps^1)$. We have derived the subleading ${\cal O}(\eps^1)$ terms by extending the results of Ref.~\cite{Czakon:2008ii}. For the manipulations involving harmonic polylogarithms \cite{Remiddi:1999ew} we have used our own software, as well as the program {\it HPL} \cite{Maitre:2007kp}. All integral convolutions are computed numerically. 

Next we consider the $q\bar q$ initiated reaction. The NLO coefficient function $s^{(1)}_{q \bar q}$ can be found in Ref.~\cite{Czakon:2008ii}. The complete NNLO cross-section $s^{(2)}_{q \bar q}$ reads:
\begin{eqnarray}
s^{(2)}_{q \bar q} &=& \tilde s^{(2)}_{q \bar q} +  \left({1\over 2\pi}\right)^2 \tilde s^{(0)}_{q \bar q} \otimes \Bigg\{ {1\over \epsilon^2}\left[ 2  P^{(0)}_{qq}\otimes P^{(0)}_{qq} + P^{(0)}_{qg}\otimes P^{(0)}_{gq} - \beta_0 P^{(0)}_{qq}\right] + {1\over \epsilon} P^{(1)}_{qq} \Bigg\} \nonumber\\
&&+ {1\over \epsilon^2} \left({1\over 2\pi}\right)^2 \tilde s^{(0)}_{gg} \otimes P^{(0)}_{gq}\otimes P^{(0)}_{gq}  + {2\over \epsilon} \left({1\over 2\pi}\right) \tilde s^{(1)}_{q \bar q} \otimes P^{(0)}_{qq} + {2\over \epsilon} \left({1\over 2\pi}\right) \tilde s^{(1)}_{q g} \otimes P^{(0)}_{gq} \, ,
\label{shat2}
\end{eqnarray}
where the NLO splitting function reads $P^{(1)}_{qq} = P^{(1),S}_{qq} +\, P^{(1),V}_{qq}$. 

We recall that in Ref.~\cite{Baernreuther:2012ws} a subset of the Double-Real diagrams corresponding to the partonic process (\ref{eq:qqbarID}) were neglected due to their small size. Despite its phenomenological insignificance, however, the reaction (\ref{eq:qqbarID}) generates collinear singularities starting from $1/\eps^{2}$ which, in Ref.~\cite{Baernreuther:2012ws}, were excluded from Eq.~(\ref{shat2}) in order to ensure consistent collinear subtraction. 

The contribution to Eq.~(\ref{shat2}) that was excluded from Ref.~\cite{Baernreuther:2012ws} reads:
\begin{eqnarray}
\Delta \tilde s^{(2)}_{q\bar q, {\rm NS}} &=& \left({1\over 2\pi}\right)^2 \tilde s^{(0)}_{q \bar q} \otimes \Bigg\{ {1\over \epsilon^2} P^{(0)}_{qg}\otimes P^{(0)}_{gq} + {1\over \epsilon} \left[ P^{(1),S}_{qq} + S^{(1)}_{qq}\right] \Bigg\} \nonumber\\
&& + {1\over \epsilon^2} \left({1\over 2\pi}\right)^2 \tilde s^{(0)}_{gg} \otimes P^{(0)}_{gq}\otimes P^{(0)}_{gq} + {2\over \epsilon} \left({1\over 2\pi}\right) \tilde s^{(1)}_{q g} \otimes P^{(0)}_{gq} \, ,
\label{collinear-special-term}
\end{eqnarray}
i.e. the above result needs to be subtracted from the RHS of Eq.~(\ref{shat2}) to arrive at the result of Ref.~\cite{Baernreuther:2012ws}.

The origin of the terms involving $P^{(0)}_{gq},~P^{(0)}_{qg}$ and $P^{(1),S}_{qq}$ in Eq.~(\ref{collinear-special-term}) is easy to understand: they involve iterated emissions that are consistent with the initial and final states of the reaction (\ref{eq:qqbarID}). The only subtle contribution to Eq.~(\ref{collinear-special-term}) is the function $S^{(1)}_{qq}$ that reads
\begin{eqnarray}
S^{(1)}_{qq} &=& \left( C_F^2 - {C_FC_A\over 2}\right) \left[ 8 - 7 x + {5-2 x^2\over 1 - x}\ln(x)  \right.\nonumber\\
&&\left.  +  {1+x^2 \over 1 - x}\left( {\pi^2\over 3} - 2 \ln(1 - x) \ln(x) + \ln^2(x) - 2 {\rm Li}_2(x) \right)\right] \, .
\label{S1qq}
\end{eqnarray}

The function $S^{(1)}_{qq}$ is a partial contribution to the space-like splitting function $P^{(1),V}_{qq}$ and originates in the interference of the splitting process $q\to q+q+\bar q$. This interference term cannot be extracted from $P^{(1),V}_{qq}$ based on its color factor $C_F(C_F-C_A/2)$ which is shared by a number of gluon emission diagrams that also contribute to $P^{(1),V}_{qq}$. We derive the function $S^{(1)}_{qq}$ with the help of two independent direct calculations, which we describe next. 

First, by extending the results of Ref.~\cite{Melnikov:2004bm}, we compute directly the time-like (fragmentation) analogue $T^{(1)}_{qq}$ of the function $S^{(1)}_{qq}$. Then, following Ref.~\cite{Mitov:2006ic}, we analytically continue $T^{(1)}_{qq}$ to space-like kinematics. For this particular contribution the analytical continuation is trivial and is just the usual replacement $f(x) \to -x f(1/x)$ supplemented by standard analytical continuation across branch points for the involved logarithmic and polylogarithmic functions. Second, we identify the function $S^{(1)}_{qq}$ as the second diagram from the class C in Fig. 7 of  Ref.~\cite{Curci:1980uw}. Since the result for this diagram is not available in that reference, we have directly computed it, following the methods of Ref.~\cite{Curci:1980uw}.  Both calculations lead to Eq.~(\ref{S1qq}).

Finally, as a by product of our calculation, we present for the first time the {\it time-like} function $T^{(1)}_{qq}$:
\begin{eqnarray}
T^{(1)}_{qq} + S^{(1)}_{qq} = \left( C_F^2 - {C_FC_A\over 2}\right) \left[ 15 (1 - x) + 7 (1 + x) \ln(x) + {1 + x^2\over 1 - x} \ln^2(x)\right] \, ,
\label{T1qq}
\end{eqnarray}
which has appeared in the literature on heavy flavor fragmentation \cite{Neubert:2007je,Ferroglia:2012ku}. 

To derive the partonic reaction (\ref{eq:qqbarID}) we use the collinear subtraction term Eq.(\ref{collinear-special-term}) 
\begin{eqnarray}
s^{(2)}_{q \bar q, {\rm NS}} = \tilde s^{(2)}_{q \bar q, {\rm NS}} + \Delta \tilde s^{(2)}_{q\bar q, {\rm NS}} \, .
\label{shat2NS}
\end{eqnarray}

Adding the result Eq.~(\ref{shat2NS}) derived in the present paper to the one derived in Ref.~\cite{Baernreuther:2012ws} we obtain the complete contribution to the $q\bar q$ initiated reaction $q\bar q\to t\bar t +X$ at NNLO. Since the contribution from the counterterm $\Delta \tilde s^{(2)}_{q\bar q, {\rm NS}}$ (\ref{collinear-special-term}) cancels in the complete $q\bar q\to t\bar t +X$ result, the point-wise cancellation of the collinear singularities (within the numerical precision) observed both in this paper and in Ref.~\cite{Baernreuther:2012ws} serves as an additional check of our setup.

\section{Scale dependence}\label{sec:scale}

The scale dependent terms $\sigma^{(1,1)}_{ij},~\sigma^{(2,1)}_{ij}$ and $\sigma^{(2,2)}_{ij}$ in Eq.~(\ref{eq:sigmapart}) can be derived from: a) the requirement that the measured hadronic cross-section $\sigma_{\rm tot}$ in Eq.~(\ref{eq:sigmatot}) be independent of the factorization scale $\mu$ through NNLO, b) the parton distribution functions $f_i$ satisfy the DGLAP evolution equations, and, c) the known running of the strong coupling constant. 

It is again natural to work in terms of the functions $s^{(n(,m))}_{ij}(\rho) \equiv \sigma^{(n(,m))}_{ij}(\rho)/\rho$:
\begin{eqnarray}
\label{scales-general}
s^{(1,1)}_{ij} &=& {1\over 2\pi} \left[ 2\beta_0 s^{(0)}_{ij} - P^{(0)}_{ki}\otimes s^{(0)}_{kj} - s^{(0)}_{ik}\otimes P^{(0)}_{kj}\right]\, , \\
s^{(2,2)}_{ij} &=& {1\over (2\pi)^2} \left[ 3\beta_0^2 s^{(0)}_{ij} - {5\over 2}\beta_0 P^{(0)}_{ki}\otimes s^{(0)}_{kj} - {5\over 2} \beta_0 s^{(0)}_{ik}\otimes P^{(0)}_{kj} \right.\nonumber\\
&&\left. + {1\over 2} P^{(0)}_{ki}\otimes P^{(0)}_{lk}\otimes s^{(0)}_{lj} 
+ {1\over 2} s^{(0)}_{il}\otimes P^{(0)}_{lk}\otimes P^{(0)}_{kj} 
+ P^{(0)}_{ki}\otimes s^{(0)}_{kl}\otimes P^{(0)}_{lj} \right] \, ,\nonumber\\
s^{(2,1)}_{ij} &=& {1\over (2\pi)^2} \left[2\beta_1 s^{(0)}_{ij}  - P^{(1)}_{ki}\otimes s^{(0)}_{kj} - s^{(0)}_{ik}\otimes P^{(1)}_{kj}\right] + {1\over 2\pi} \left[ 3\beta_0 s^{(1)}_{ij}  - P^{(0)}_{ki}\otimes s^{(1)}_{kj} - s^{(1)}_{ik}\otimes P^{(0)}_{kj}\right]\, .\nonumber
\end{eqnarray}

The powers of $1/(2\pi)$ appearing in the above equations originate in the somewhat unconventional choice of $\as^n$ as the expansion parameter in Eq.~(\ref{eq:sigmapart}). The expansion of the splitting functions is as in Eq.~(\ref{Gamma}) where $\beta_0$ is also defined. The two-loop beta-function coefficient reads $\beta_1 = 17 C_A^2/6 - 5 C_A N_L/6 - C_F N_L/2$.

The scale dependence for any specific reaction can be easily derived from the above equations. The expression for the $q\bar q$ reaction has been given in Ref.~\cite{arXiv:0906.5273}. The scale-dependent terms for the reaction $q\tilde q$ are not available in the literature and we give them here:
\begin{eqnarray}
s^{(2,2)}_{q\tilde q} &=& {1\over (2\pi)^2} \left[ 
s^{(0)}_{q\bar q}\otimes P^{(0)}_{qg}\otimes P^{(0)}_{gq} + s^{(0)}_{gg}\otimes P^{(0)}_{gq}\otimes P^{(0)}_{gq}\right] \, ,\nonumber\\
s^{(2,1)}_{q\tilde q} &=& -{2\over (2\pi)^2}~ s^{(0)}_{q\bar q}\otimes P^{(1)}_{q\overline{\tilde q}} - {2\over 2\pi} ~ s^{(1)}_{gq}\otimes P^{(0)}_{gq}\, ,
\label{scales-qtildeq}
\end{eqnarray}
where the splitting function $P^{(1)}_{q\overline{\tilde q}}$ is given in Eq.~(\ref{P1qq}) and $s^{(n)}_{ij}, ~n=0,1$ are the finite LO and NLO coefficient functions available in analytical form \cite{Czakon:2008ii}.

We have computed all convolutions numerically and produced our own fits for all scaling functions. We have implemented them in the program {\tt Top++} \cite{Czakon:2011xx}: the ones for the complete $q\bar q$ reaction in version 1.2 and the ones for the $qq,~qq'$ and $q\bar q'$ reactions in version 1.3.

\section{Results}\label{sec:results}

We calculate the coefficient functions $ \sigma^{(2)}_{ij}$ for the reactions (\ref{eq:qqbarID},\ref{eq:qqbar'},\ref{eq:qq'},\ref{eq:qq}) numerically in a number of points on the interval $\beta\in (0,1)$. For short, we will sometimes refer to the set of computed points and their numerical uncertainties as ``data". Specifically, the $qq'$ and $q\bar q'$ initiated reactions are computed in 80 points, with $\beta_{80}=0.999$, as was also done in Ref.~\cite{Baernreuther:2012ws}. For the $q\bar q$ initiated reaction (\ref{eq:qqbarID}) we have added the point $\beta = 0.9999$, while for the $qq$-initiated reaction (\ref{eq:qq}) we have added two more points $\beta = 0.99375$ and $\beta = 0.9999$. 

The reason for including these additional points is to more accurately constrain the high-energy $\beta \to 1$ behavior of the numerically extracted partonic cross-sections. As is well known, the partonic reactions considered in this paper exhibit logarithmic rise at high energy due to diagrams where the top-pair is emitted in the $t$-channel. The leading behavior in the limit $\beta \to 1$ (or, equivalently, in the limit $\rho \to 0$) of the partonic cross-sections for all four reactions (\ref{eq:qqbarID},\ref{eq:qqbar'},\ref{eq:qq'},\ref{eq:qq}) is
\begin{equation}
\sigma^{(2)}_{f_1f_2\to t\bar t f_1f_2} \Big\vert_{\rho \to 0} \approx c_1\ln(\rho) + c_0 +{\cal O}(\rho) \, .
\label{eq:high-ebergy-limit}
\end{equation}
The constant $c_1$ has been predicted exactly in Ref.~\cite{Ball:2001pq}. Its numerical value is 
\begin{equation}
c_1 = - 0.4768323995789214 \, .
\label{eq:c1}
\end{equation}

We have verified that for all four reactions (\ref{eq:qqbarID},\ref{eq:qqbar'},\ref{eq:qq'},\ref{eq:qq}) our numerical calculations (with unconstraint fits) return values for $c_1$ that are within $2\%$ from the exact  result (\ref{eq:c1}). Therefore, having verified the consistency of our calculation with the exactly predicted leading logarithmic term, in all subsequent fits we impose the exact value for the leading logarithmic term. This allows us to extract the constant $c_0$ with maximum precision, which turns out to be high enough to significantly improve the approximate prediction that has recently appeared in the literature \cite{Moch:2012mk}, and to derive fits that are highly accurate even in the limit $\beta \to 1$. 

As we already anticipated in Ref.~\cite{Baernreuther:2012ws}, and confirm in this paper, the contributions from the all-fermionic reactions to the total inclusive top-pair production cross-section at present hadron colliders are negligible. A more detailed analysis will be performed in section \ref{sec:pheno}.

\subsection{$qq'$ and $q\bar q'$ initiated reactions.}\label{sec:prime-rections}

\begin{figure}[tbp]
\centering
\includegraphics[width=.95\textwidth]{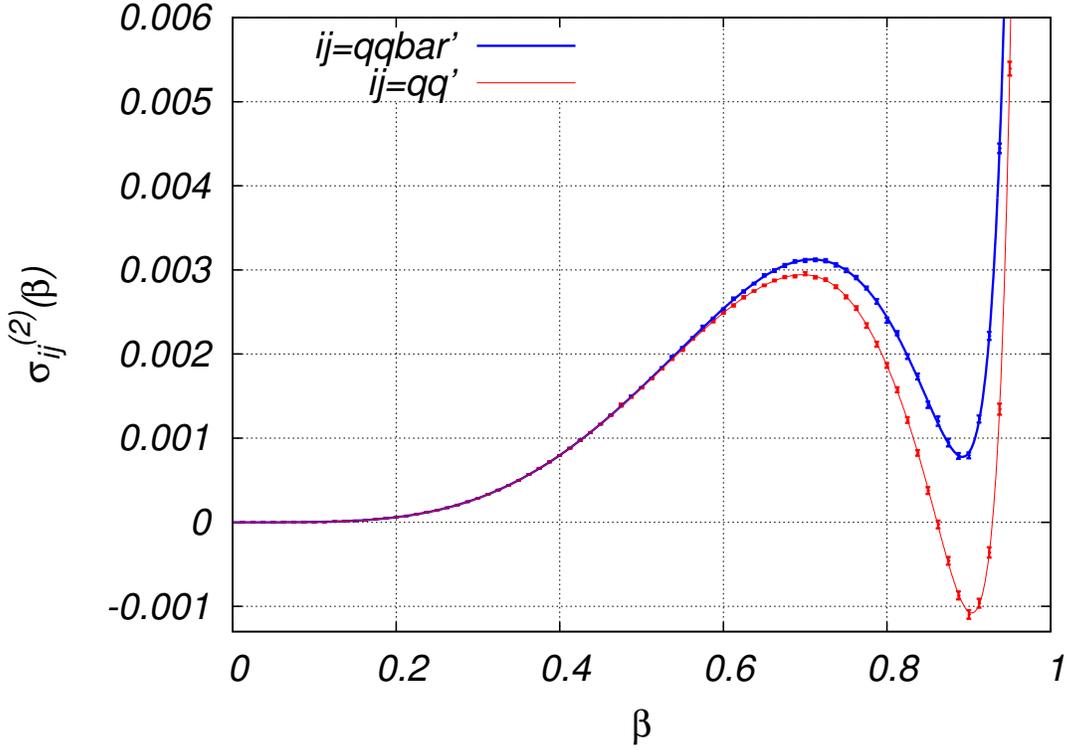}
\caption{\label{fig:qqbarp-qqp} The computed results, including numerical uncertainties, for the partonic cross-sections $\sigma^{(2)}_{q\bar q'}$ (blue) and $ \sigma^{(2)}_{qq'}$ (red). The discrete results, computed in 80 points, are overlaid with the corresponding analytical fits (see text). Both cross-sections diverge logarithmically in the limit $\beta\to 1$.}
\end{figure}

The results for the partonic cross-sections for these two reactions read:
\begin{eqnarray}
\label{eq:fit20qqbar'}
\sigma^{(2)}_{q\bar q'} &=& c_1 \lrho-\beta^2 \exp\left(f_{q\bar q'}\right) \, , \\
f_{q\bar q'} &=& -0.740572-31.2117 \beta^2-0.31495 \beta^3+15.8601 \beta^4-1.64639 \beta^5+18.9767 \beta^6\nonumber\\
&& +\lrhosq \left(-3.16565 \rho+12.3828 \rho^2\right)+\lrho \left(-19.6977 \rho-16.1386 \rho^2+4.17707 \rho^3\right) \, , \nonumber\\
&&\nonumber\\
\label{eq:fit20qq'}
\sigma^{(2)}_{qq'} &=& c_1 \lrho-\beta^2 \exp\left(f_{qq'}\right) \, , \\
f_{qq'} &=& -0.740558-23.4518 \beta^2-0.193073 \beta^3-5.97215 \beta^4-0.541402 \beta^5+31.8227 \beta^6\nonumber\\
&& + \lrhosq \left(-3.29162 \rho+15.9932 \rho^2\right)+\lrho \left(-21.3725 \rho-11.1642 \rho^2+8.64746 \rho^3\right) \, . \nonumber
\end{eqnarray}
The constant $c_1$ is given in Eq.~(\ref{eq:c1}). The analytical expressions in Eqs.~(\ref{eq:fit20qqbar'},\ref{eq:fit20qq'}) are derived as global fits of the set of 80 points we compute numerically. The data, and the corresponding fits, are plotted on Fig.~\ref{fig:qqbarp-qqp}. 

As is evident from Fig.~\ref{fig:qqbarp-qqp} these partonic cross-sections vanish at threshold $\beta=0$ and diverge logarithmically in the high-energy limit $\beta \to 1$. The quality of the fits (\ref{eq:fit20qqbar'},\ref{eq:fit20qq'}) is very high for intermediate and large values of $\beta$, i.e. in this region the precision of the results is restricted by the numerical precision of our numerical evaluation. Fitting in the region of small $\beta$ turns out to be more problematic, however, since the two functions are as small as ${\cal O}(10^{-10})$ over a sizable range of $\beta$. In this range the distance between the fits (\ref{eq:fit20qqbar'},\ref{eq:fit20qq'}) and the data is large compared to the size of the numerical uncertainty. However, the absolute size of the deviation $data - fit$ is below ${\cal O}(10^{-6})$, i.e. the inaccuracy of the fits is completely immaterial for any foreseeable phenomenological application of these results. We find the simplicity of the analytical expressions in Eqs.~(\ref{eq:fit20qqbar'},\ref{eq:fit20qq'}) very appealing. Because of their very high absolute accuracy we have implemented them in the program {\tt Top++} \cite{Czakon:2011xx}.

From the fits we extract the following values for the constant $c_0$:
\begin{equation}
c_0 ~ ({\rm from~Eqs.}~ (\ref{eq:fit20qqbar'},\ref{eq:fit20qq'})) = \left\{ 
  \begin{array}{l l}
    	-2.5173 & ~ {\rm from}\quad \sigma^{(2)}_{q\bar q'} \, , \\
   	-2.5186 & ~ {\rm from}\quad \sigma^{(2)}_{qq'} \, . \\
  \end{array} \right.
\label{c0basic}
\end{equation}
We note that the values of $c_0$ extracted from both reactions are compatible within the numerical uncertainty. In the following we turn to the estimation of the uncertainty in the extracted value for $c_0$.

Due to the global nature of the fitting procedure, one might wonder how the quality of the fit at low $\beta$ affects the quality of the fits in the phenomenologically more relevant region of large $\beta$ and, in particular, the extraction of the divergent behavior at $\beta \to 1$.
To estimate the robustness of the extracted high-energy behavior of the partonic cross-sections we derive second set of fits, with the aim of fitting both the small- and large-$\beta$ regions. These alternative fits fit the data within the numerical uncertainties, except for the first few points, where they deviate from data within about 10 times the size of the numerical error. The absolute size of this deviation is ${\cal O}(10^{-10})$. The analytical form of these fits is much more cumbersome and the values of their parameters are highly tuned. For this reason we do not present this set of fits explicitly. Moreover, the phenomenological implication of the differences between the two sets of fits is completely immaterial.

The only place where the difference between the two rather extreme fits plays a role is in the very large $\beta$ behavior of the partonic cross-sections and in the extraction of the constant $c_0$. Indeed, from this alternative set of fits we obtain
\begin{equation}
c_0 ({\rm alternative~fits}) = \left\{ 
  \begin{array}{l l}
    	-2.4134 & ~ {\rm from}\quad \sigma^{(2)}_{q\bar q'} \, , \\
   	-2.4037 & ~ {\rm from}\quad \sigma^{(2)}_{qq'} \, , \\
  \end{array} \right.
\label{c0alternativefits}
\end{equation}
Again, the extracted values of $c_0$ from the two reactions are compatible. We take the difference between the two types of fits, Eq.~(\ref{c0basic}) and Eq.~(\ref{c0alternativefits}), as a measure of the uncertainty in the extraction of the constant $c_0$ from our calculation, which we estimate around $5\%$.

Next we compare our result for the constant $c_0$ with the corresponding prediction of Ref.~\cite{Moch:2012mk}. The value for $c_0$ predicted in Ref.~\cite{Moch:2012mk} has substantial uncertainty, slightly above $50\%$, and is predicted in the range $(-1.4305,-2.43185)$. We see that our value for $c_0$ is consistent with the prediction of Ref.~\cite{Moch:2012mk}, albeit at the end of the uncertainty range quoted in that reference, and has an order of magnitude better precision. Overall, the agreement we find with the prediction of Ref.~\cite{Moch:2012mk} (which was derived with completely different methods) is a non-trivial check for both setups.

Finally, we would like to point out that the knowledge of the high-energy behavior (\ref{eq:high-ebergy-limit}) of the partonic cross-sections alone is insufficient for meaningful collider phenomenology. The reason for this is that the high-energy expansion of the partonic cross-sections is not well converging and thus not a good approximation outside the range of $\beta\approx 1$; it is only relevant for the description of heavy pair production at very large $\beta$ which is not the case for top-pair production at the Tevatron and LHC. 
\begin{figure}[tbp]
\centering
\includegraphics[width=.95\textwidth]{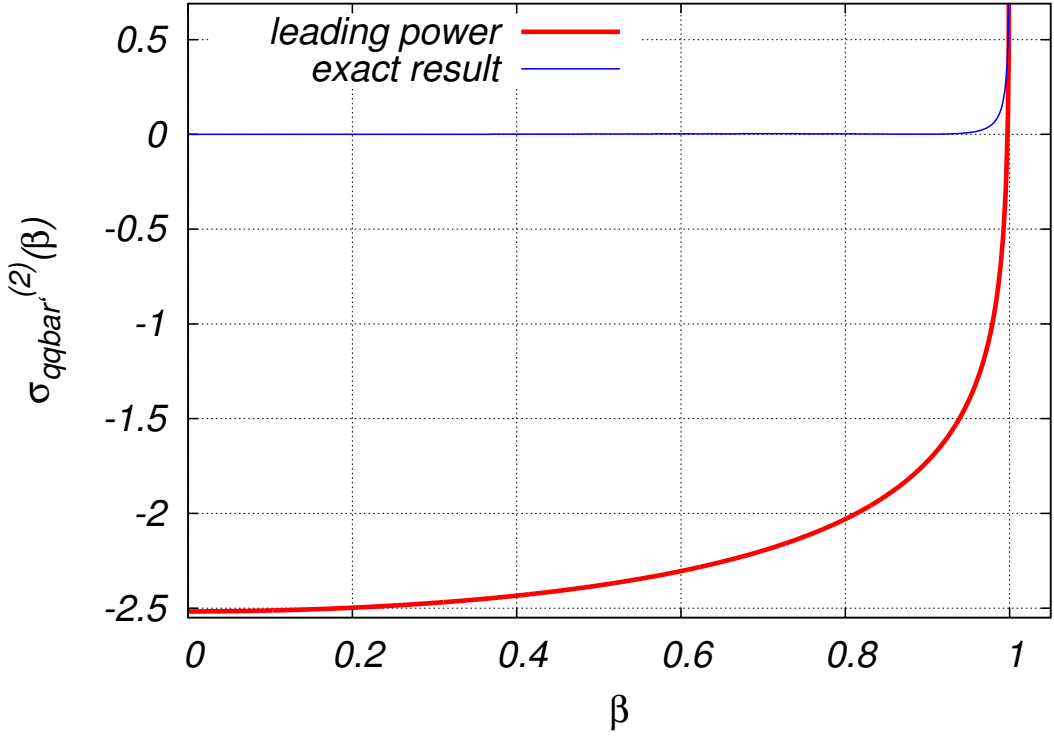}
\caption{\label{fig:qqbarp-leading-power} Comparison of the partonic cross-sections $\sigma^{(2)}_{q\bar q'}$ (blue) and its leading power behavior in the high-energy limit Eq.(\ref{eq:high-ebergy-limit},\ref{c0basic}) (red). The blue curve, same as the blue cure on Fig.~\ref{fig:qqbarp-qqp}, appears to be zero on the scale of the red curve outside the narrow range $\beta\approx 1$.}
\end{figure}

To visualize this better, on Fig.~\ref{fig:qqbarp-leading-power} we plot the cross-section $\sigma^{(2)}_{q\bar q'}$ and its high-energy leading-power approximation (\ref{eq:high-ebergy-limit},\ref{c0basic}). It is easy to see that the behavior of the two functions is dramatically different, by three orders of magnitude or more, outside the narrow range $\beta\approx 1$. We note that Fig.~\ref{fig:qqbarp-leading-power} looks similarly for any one of the reactions (\ref{eq:qqbarID},\ref{eq:qqbar'},\ref{eq:qq'},\ref{eq:qq}).

\subsection{$qq$ and $q\bar q ({\rm NS})$ initiated reactions.}

\begin{figure}[tbp]
\centering
\includegraphics[width=.95\textwidth]{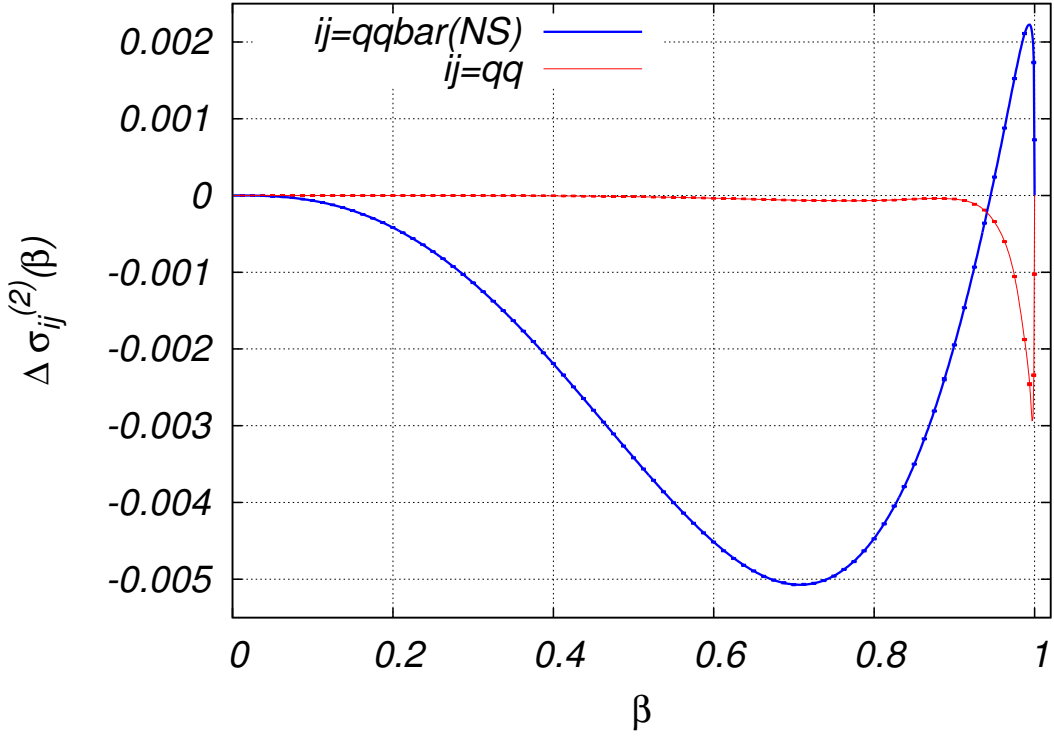}
\caption{\label{fig:qqbar-qq} The computed results, including numerical uncertainties, for the interference partonic cross-sections $\Delta\sigma^{(2)}_{q\bar q,{\rm NS}}$ (blue) and $\Delta\sigma^{(2)}_{qq}$ (red). The discrete results, computed in respectively 81 and 82 points, are overlaid with the corresponding analytical fits (see text). Both results vanish in the limit $\beta\to 1$.}
\end{figure}

As emphasized in section \ref{sec:calc}, we compute the contributions to the reactions $qq$ and $q\bar q ({\rm NS})$ as differences with respect to, respectively, the $qq'$ and $q\bar q'$ processes. These differences, $\Delta \sigma^{(2)}_{q\bar q, {\rm NS}}$ and $\Delta\sigma^{(2)}_{qq}\equiv \sigma^{(2)}_{qq}-\sigma^{(2)}_{qq'}$, vanish in both the threshold and high-energy limits and read:
\begin{eqnarray}
\label{eq:fit20qqbarNS}
\sigma^{(2)}_{q\bar q, {\rm NS}} &=& \sigma^{(2)}_{q\bar q'} + \Delta \sigma^{(2)}_{q\bar q, {\rm NS}} \, , \\
\Delta \sigma^{(2)}_{q\bar q, {\rm NS}} &=& \left(1.53647 \beta^3 + 10.7411 \beta^4\right) \rho - 24.3298 \beta^4 \rho^2 + \left(-4.50719 \beta^3 + 15.4975 \beta^4\right) \rho^3 \nonumber\\
&& + \left(2.90068 \beta^3 - 4.98808 \beta^4\right) \rho^4 -1.26644 \beta^{20} \lbe \nonumber\\
&& + \lrhosq \left(0.327143 \rho - 10.7669 \rho^2\right) + \lrho \left(3.86236 \rho - 21.332 \rho^2 + 17.4705 \rho^3\right)   \, , \nonumber\\
&&\nonumber\\
\label{eq:fit20qq}
\sigma^{(2)}_{qq} &=& c_1 \lrho-\beta^2 \exp\left(f_{qq}\right) \, , \\
f_{qq} &=& -0.740558 - 22.8129 \beta^2 - 0.191648 \beta^3 - 6.58031 \beta^4 - 0.537669 \beta^5 + 31.7872 \beta^6 \nonumber\\
&&  +  \lrhosq \left(-3.25313 \rho + 15.8988 \rho^2\right) + \lrho \left(-21.0783 \rho - 10.8176 \rho^2 + 8.64557 \rho^3\right) \, . \nonumber
\end{eqnarray}
The constant $c_1$ is defined in Eq.~(\ref{eq:c1}). The data and the fits for the functions $\Delta \sigma^{(2)}_{q\bar q, {\rm NS}}$ and $\Delta\sigma^{(2)}_{qq}$ are plotted on Fig.~\ref{fig:qqbar-qq}.

The results for the functions $\Delta\sigma^{(2)}_{q\bar q,{\rm NS}}$ and $\Delta\sigma^{(2)}_{qq}$ demonstrate that these functions vanish in both limits $\beta = 0,1$. Therefore, the high-energy behavior of the complete reactions $\sigma^{(2)}_{q\bar q,{\rm NS}}$ and $\sigma^{(2)}_{qq}$ is determined by the functions $\sigma^{(2)}_{q\bar q'}$ and $\sigma^{(2)}_{qq'}$ discussed in detail in section \ref{sec:prime-rections}.

The quality of the fits in Eqs.~(\ref{eq:fit20qqbarNS},\ref{eq:fit20qq}) is similar to the ones in section \ref{sec:prime-rections}. The absolute quality of Eq.~(\ref{eq:fit20qqbarNS}) is quite good, with absolute difference $data-fit$ below ${\cal O}(10^{-7})$ for small $\beta$. The quality of the fit of $\Delta \sigma^{(2)}_{q\bar q, {\rm NS}}$ beyond the first seven lowest-$\beta$ points is dominated by the uncertainty in the numerical evaluation.

For the $qq$-initiated reaction we have performed two fits. The fit in Eq.~(\ref{eq:fit20qq}) is performed for the total contribution $\sigma^{(2)}_{qq}$. Its absolute quality is also high, with absolute difference $data-fit$ below ${\cal O}(10^{-8})$ for small $\beta$. The quality of the fit of $\sigma^{(2)}_{qq}$ beyond the first ten lowest-$\beta$ points is dominated by the uncertainty in the numerical evaluation. From this fit we extract the following value for the constant $c_0$:
\begin{equation}
c_0 ~ ({\rm from~Eq.}~ (\ref{eq:fit20qq})) = -2.5196  \, ,
\label{c0basicqq}
\end{equation}
which is consistent with the ones extracted from Eqs.~(\ref{eq:fit20qqbar'},\ref{eq:fit20qq'}).

Finally we have performed a tighter, higher quality fit for the difference $\Delta\sigma^{(2)}_{qq}$. We do not present it explicitly here for the same reasons explained in section \ref{sec:prime-rections}.

\section{Discussion}\label{sec:pheno}

In the present paper we calculate the NNLO corrections to total inclusive top-pair production at hadron colliders from the six-fermion partonic reactions (\ref{eq:qqbarID},\ref{eq:qqbar'},\ref{eq:qq'},\ref{eq:qq}). The results in this work, in particular, complete the calculation of the NNLO correction to the reaction $q\bar q\to t\bar t + X$ \cite{Baernreuther:2012ws}. The contributions from these reactions have been discussed in the recent literature \cite{Moch:2012mk,Beneke:2012wb,Schilling:2012dx}.

As we already anticipated in Ref.~\cite{Baernreuther:2012ws}, and confirm with our present calculation, the contributions from the all-fermionic reactions are phenomenologically insignificant for top-pair production at present hadron colliders like Tevatron and LHC. The numerical contribution of all four reactions (\ref{eq:qqbarID},\ref{eq:qqbar'},\ref{eq:qq'},\ref{eq:qq}) to the top-pair production cross-section at the Tevatron and LHC is presented in table \ref{tab:table}.
\begin{table}[ht]
\begin{center}
\begin{tabular}{| c | c | c | c | c |}
\hline
& {\rm Tevatron}  & {\rm LHC}~7~{\rm TeV} &  {\rm LHC}~8~{\rm TeV} & {\rm LHC}~14~{\rm TeV} \\
\hline 
$\Delta\sigma_{q\bar q,({\rm NS})}$~[pb] & -0.0020 & -0.0097 & -0.0124 & -0.0299 \\ 
\hline 
$\sigma_{q\bar q,({\rm NS})}$~[pb] & -0.0009 & -0.0001 & 0.0021 & 0.0464 \\ 
\hline 
$\sigma_{{\rm all}}$~[pb] & 0.0003 & 0.0970 & 0.1504 & 0.7885 \\ 
\hline
$\sigma_{{\rm tot}}$~[pb] & 7.0056 & 154.779 & 220.761 & 852.177 \\ 
\hline
\end{tabular}
\caption{\label{tab:table} Contribution to the total top-pair inclusive cross-section due to the reactions computed in this work: due to the reaction (\ref{eq:qqbarID}) alone, $\Delta\sigma_{q\bar q,({\rm NS})}$ and $\sigma_{q\bar q,({\rm NS})}$, and due to all four reactions (\ref{eq:qqbarID},\ref{eq:qqbar'},\ref{eq:qq'},\ref{eq:qq}) combined, $\sigma_{{\rm all}}$. As a reference point, our pure fixed order prediction for $\sigma_{{\rm tot}}$ is also given.}
\end{center}
\end{table}
Specifically, we present separately the results for $\Delta\sigma_{q\bar q,({\rm NS})}$ and $\sigma_{q\bar q,({\rm NS})}$ due to the reaction (\ref{eq:qqbarID}) as well as the combined effect $\sigma_{{\rm all}}$ due to all four reactions considered in this paper. As a point of reference we also present  in table~\ref{tab:table}  the pure fixed order NNLO prediction $\sigma_{{\rm tot}}$ for the total inclusive cross-section. 

The contributions from the reactions (\ref{eq:qqbarID},\ref{eq:qqbar'},\ref{eq:qq'},\ref{eq:qq})  are in the sub-permil range, both for central values and scale variation, for Tevatron and LHC at 7,8 and 14 TeV. The numbers in table~\ref{tab:table} are computed in fixed order QCD with version 1.3 of the program {\tt Top++} \cite{Czakon:2011xx} with default precision, $m_t=173.3~{\rm GeV}$, central scales and MSTW2008nnlo68cl pdf set~\cite{Martin:2009iq}.

The results of the present paper might potentially be of interest for the description of lighter quark production ($b$ or $c$) or for top-pair production at possible future high-energy hadron colliders. Only in such cases, due to the partonic flux being peaked towards larger values of $\beta$, the high-energy rise of the reactions (\ref{eq:qqbarID},\ref{eq:qqbar'},\ref{eq:qq'},\ref{eq:qq}) might become phenomenologically relevant. 

We derive high-quality analytical fits for the partonic cross-sections in all four reactions (\ref{eq:qqbarID},\ref{eq:qqbar'},\ref{eq:qq'},\ref{eq:qq}). Our fits have the exact leading logarithmic behavior \cite{Ball:2001pq} in the high-energy limit. Therefore, any numerical difference in this limit due to the imprecision of our fits behaves no worse than a constant at large $\beta$, i.e. as $c_0^{\rm exact} - c_0^{\rm fit} + {\cal O}(\rho)$. Based on our findings in section \ref{sec:prime-rections} we estimate
\begin{equation}
c_0^{\rm exact} - c_0^{\rm fit} \leq {\cal O}(10^{-1})\, .
\label{eq:c0-estimate}
\end{equation}

On the other hand, up to the point $\beta_{80}=0.999$, our fits are quite accurate, typically much better than $1\%$, and thus a very good representation of the exact result. Therefore, it is only in the region beyond the point $\beta_{80}=0.999$ where the difference (\ref{eq:c0-estimate}) might start accumulating error. Barring extreme cases, however, we believe that all NNLO partonic cross-section fits derived by us so far are under good theoretical control in the full kinematical range. 

Finally, we would like to stress that the knowledge of the high-energy behavior (\ref{eq:high-ebergy-limit}) of the partonic cross-sections alone is insufficient for meaningful phenomenology. The reason for this is that the high-energy expansion of the partonic cross-sections is not well converging; see Fig.~\ref{fig:qqbarp-leading-power}.  The high-energy expansion is only relevant for the description of heavy pair production at very large $\beta$ which is not the case of top-pair production at the Tevatron and LHC.

Work on the calculation of the NNLO corrections to the two remaining partonic reactions $qg\to t\bar t+X$ and $gg\to t\bar t+X$ is ongoing and will be presented in a forthcoming publication.

\acknowledgments
The work of M.C. was supported by the Heisenberg and by the Gottfried Wilhelm Leibniz programmes of the Deutsche Forschungsgemeinschaft, and by the DFG Sonderforschungsbereich/Transregio 9 ÒComputergest\"utzte Theoretische TeilchenphysikÓ.

\end{document}